\documentclass{article}

\usepackage{arxiv}

\usepackage[utf8]{inputenc} 
\usepackage[T1]{fontenc}    
\usepackage{hyperref}       
\usepackage{url}            
\usepackage{booktabs}       
\usepackage{amsfonts}       
\usepackage{nicefrac}       
\usepackage{microtype}      
\usepackage{lipsum}		
\usepackage{graphicx}
\usepackage{natbib}
\usepackage{doi}

\title{Overview and study of the 3D-TSV interconnects induced coupling in CMOS circuits}

\author{ \href{https://orcid.org/0000-0003-0032-9845}{\includegraphics[scale=0.06]{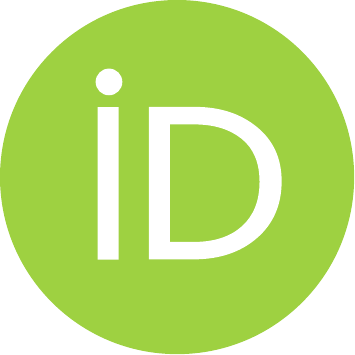}\hspace{1mm}Mohamed El Amine Benkechkache}\thanks{Correspondng author} \\
	Department of Nuclear Engineering\\
	  The University of Tennessee\\
	Knoxville, TN, USA \\
	\texttt{mbenkech@utk.edu} \\
	\And
	{\hspace{1mm}Saida Latreche} \\
	Department of Electronics\\
	The University of Constantine\\
	Constantine, Algeria \\
	 \\
  	\And
	{\hspace{1mm}Lamis Ghoualmi} \\
	Department of Computer Science\\
	The University of Constantine\\
	Constantine, Algeria \\
	 \\
}

\renewcommand{\shorttitle}{TSV interconnects with CMOS ICs}

\hypersetup{
pdftitle={Pre-print},
pdfsubject={q-bio.NC, q-bio.QM},
pdfauthor={Mohamed El Amine Benkechkache, Elias D.~Striatum},
pdfkeywords={First keyword, Second keyword, More},
}

\begin{document}
\maketitle

\begin{abstract}
The semiconductor industry's rapid advancement pushes conventional two-dimensional technology to its utmost limitations in terms of scaling, performance, and cost factors. These challenges drive the usage of 3D technology in the production of various Integrated Circuits. One of the numerous features of 3D Integration is the use of Through Silicon Vias (TSVs) for the assembly of multilayers into a single stack. This process, which was initially developed for memory chips, has been used afterward in many applications in other areas of microelectronics. The purpose of this research is to assess the effect of 3D-TSV interconnection on the performance of MOS transistors and CMOS circuits. This is accomplished using numerical and analytical models capable of describing the substrate coupling induced by TSV at the circuit level. The analytical approach proposed enables the study and optimization of the performance, not only of MOS devices but also large CMOS circuits with 3D interconnects as a function of various technological and electrical parameters.

\end{abstract}

\keywords{3D-TSV \and Analytical model \and CMOS circuits \and MOS transistors \and TCAD simulations}

\section{Introduction}

The functionality of computational systems has increased dramatically in the microelectronics industry over the last few decades. Many advancements (like speed, power, and integration level) have been driven by semiconductor device scaling, which has increased the number of transistors on a single chip in accordance with Moore's Law \cite{Pen}. Furthermore, as the semiconductor industry has evolved, the concept of "More Moore" has incorporated a variety of functionalities in the same technological node, allowing for an increase in integration density \cite{Lau,Rous}. The "More than Moore" approach, on the other hand, has focused on functional diversification, presenting a new category of devices with heterogeneous functionalities that do not scale in the same technological node. As a result, over the last decade, conventional 2D technology has been shown to have some issues, particularly with the increase in demand for device and interconnects capabilities in terms of chip size, density, minimum metal pitch, and the number of metal layers. Because of all of these constraints, the development of an alternative technology for designing and building microelectronic systems, such as 3D integration, became critical. 

3D integration technology, which was first used in memory chips, has since found its use in other domains of microelectronics, like image sensors and MEMS \cite{TNS,LISe,dalla,Lodola,Ratti}. The use of Through Silicon Vias (TSVs), which are regarded as the enabling technology for multilayer Integrated Circuits (ICs) assembled into a single package, is one of the various approaches to such technology \cite{Fen}. The chip packaging with 3D-TSV technology has resulted in numerous benefits and advantages \cite{Pap,Duko}. It enabled, in particular, to reduce power dissipation, the number of long interconnects, and their overall length, thereby increasing the speed and performance of MOS devices \cite{Lu}. However, this type of interconnects shows some undesirable effects on the performance of ICs nearby, like the substrate coupling effect or phenomena \cite{Rous,Vachan,Zhang}. Such a parasitic effect has been studied experimentally and analytically in traditional 2D circuits \cite{Shreeve,van}. This effect has been investigated experimentally and analytically in conventional 2D circuits, however, in 3D-ICs it still has yet to be investigated. Some studies concentrated on TSV electrical characterization \cite{Dong,Brocard,Jong}, stress \cite{Moon}, or thermal \cite{Chuk}, but the TSVs-induced coupling on CMOS circuits was rarely investigated. Despite the fact that the effect of TSVs interconnects on MOS devices was thoroughly studied in \cite{amine,Rousseau,Abouelatta}, the TSVs-induced coupling on complex CMOS circuits was not taken into account.

This paper is structured as follows: Section II will first evaluate the impact of 3D-TSV interconnects on MOS devices using a numerical approach as well as an analytical model utilized to describe an equivalent circuit capable of evaluating the 3D-TSV interconnects. The analytical model will then be used in sections III and IV to investigate the effect of 3D-TSV technology in CMOS circuits, specifically a CMOS inverter and an oscillator, with the goal of optimizing their performance in terms of various technological and electrical parameters. Section V summarizes the main findings.

\section{3D-TSV influence on MOS devices}

One of the major concerns of the presence of TSV interconnects nearby MOS components and CMOS circuits is the induced parasitic effects known as "the substrate coupling" which affects their electrical performance. Therefore, in order to study and evaluate such an impact, a numerical approach is considered by testing MOS components placed close to a via using TCAD simulations first. However, once considering evaluating the 3D-TSV technology on CMOS circuits, numerical simulations are time-consuming. To this purpose, we are going to define an analytical approach to study the parasitic substrate coupling induced by the via on the electrical behavior of CMOS circuits. The main objective of both numerical and analytical approaches considered is to reduce the substrate coupling induced by the presence of this kind of vertical interconnections on electronic devices in 3D technology.

\subsection{Numerical model}

To study the impact of TSV interconnections on the performance of MOS devices, a numerical model is used in this case.  

\subsubsection{Device description and analysis method}
\begin{figure}[!t]
	\centering
	\includegraphics[width=0.55\textwidth, height=0.3\textheight]{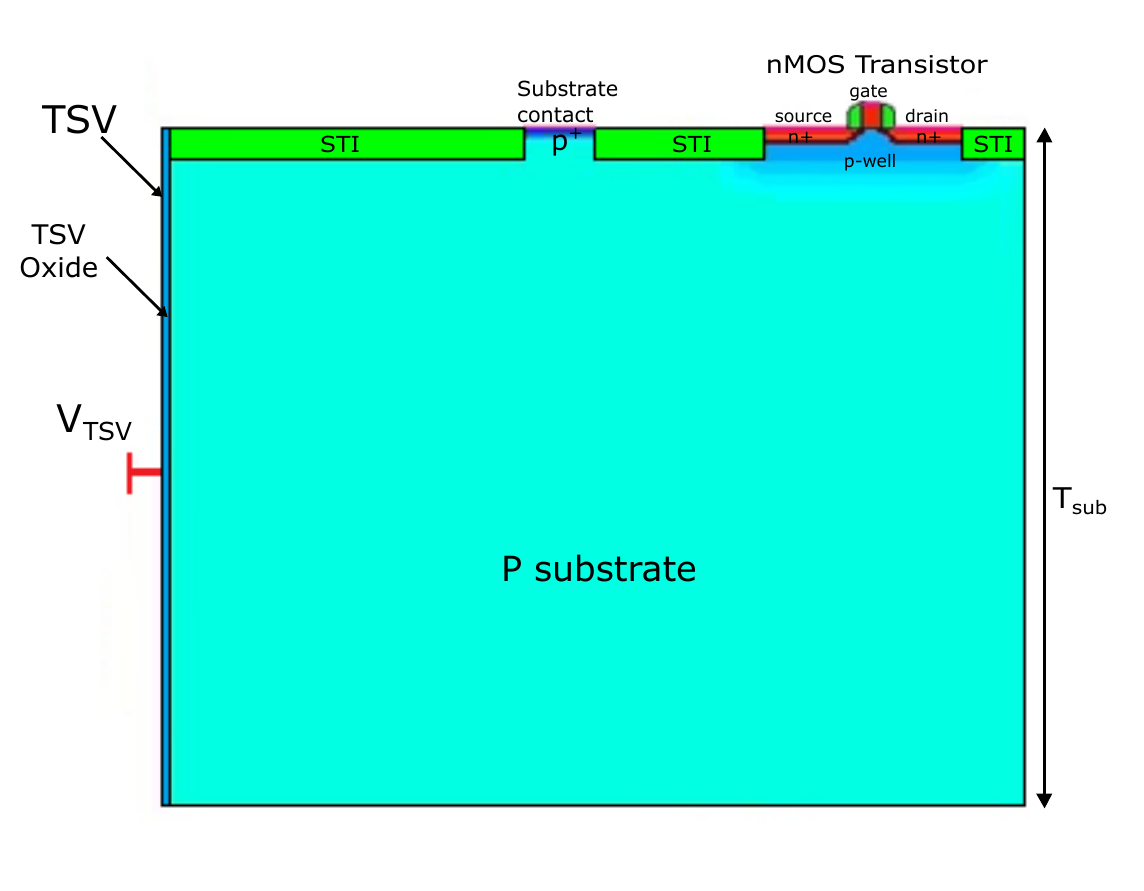}
	\caption{Two-dimensional structure showing a single nMOS transistor next to a TSV contact.}
	\label{fig:fig1}
\end{figure}

The analysis method conducted is based on numerical simulations carried out using the Sentaurus-TCAD tool. A simplified two-dimensional structure of a MOS transistor placed next to a TSV was considered as demonstrated in Figure \ref{fig:fig1}. However, since the dimensions of the TSV are much larger than those of the MOS transistor, the meshing of the assembly is difficult due to the limited number of mesh nodes. Therefore, the TSV is considered as a perfect conductor and is represented on the structure as an electrode applied on the entire left vertical boundary of the structure.

\begin{figure}[!t]
	\centering
	\includegraphics[width=1\textwidth, height=0.3\textheight]{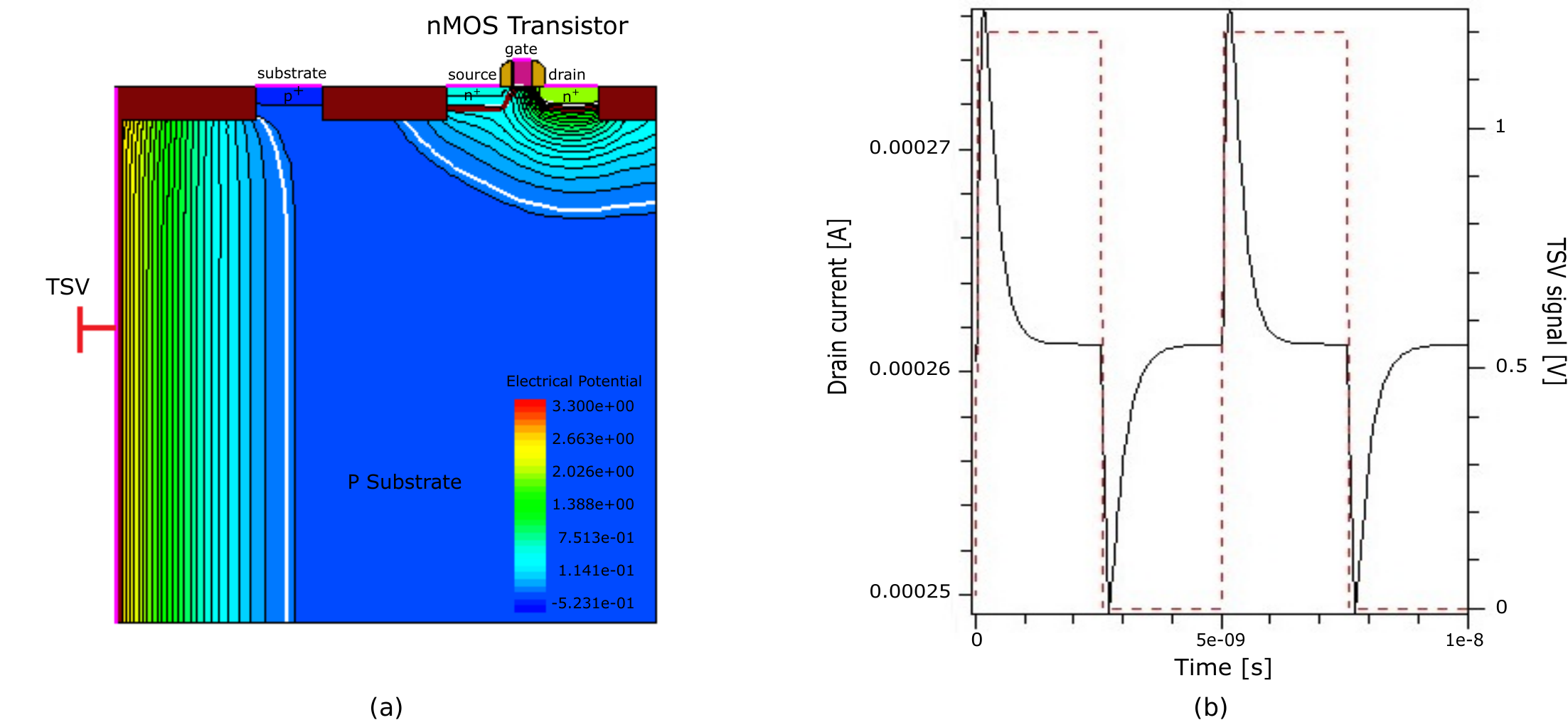}
	\caption{(a) Electrical potential distribution cross-section, and (b) impact of TSV-induced coupling on the drain current of the MOS transistor. 
		\label{fig:fig2}}
\end{figure}

The study of the substrate coupling induced by the via was carried out with a transient simulation in order to evaluate the variations in the electrical characteristics of the MOS components. Therefore, it is important initially to put the MOS transistor used in a saturation regime V$_{GS}$ = V$_{DS}$ = 1.5 V, then apply a parasitic potential on the TSV with a square signal of f = 200 MHz frequency and a maximum amplitude of 1.2 V. The induced variation in the drain current as a function of different technological and electrical parameters will be investigated. Figure \ref{fig:fig2}-a shows the potential lines propagating in the silicon substrate when applying a voltage on the via nearby.

The influence of the induced coupling by the TSV on the electrical performance of the MOS transistor results in a significant variation observed in the drain current, as shown in Figure \ref{fig:fig2}-b. These variations follow a capacitive behavior where each rising or falling edge of the potential applied to the TSV, a charging and a discharging phenomenon is noted in the behavior of the drain current. This methodology is applied to all the simulations described here. The objective of this approach is to test different geometries and configurations of the 3D architecture. 

\subsubsection{Parameter optimization}

The optimization of different parameters concerning the substrate thickness, the separation distance from the TSV, the rising/falling time of signal applied to the TSV, and the insulation thickness of the via was performed using TCAD simulations.

\begin{itemize}
\item
\textbf{Substrate thickness:}
The thickness of the substrate, or in other words the length of the TSV, is one of the most important technological parameters with regard to 3D integration. This parameter directly imposes the diameter of the vias according to the conditions of the possible form factors for the TSVs. In our case study, we have considered a range of values that corresponds to the real technological process considered for high-density 3D integration [8]. This range varies from 5 $\mu$m up to 20 $\mu$m thickness. This limiting value is imposed by the large number of meshes that would then be necessary. By considering the same transient study presented previously, the variations observed in the drain current of the MOS transistor are noted and presented in Figure \ref{fig:fig3}. This MOS transistor is placed at a distance of 6 $\mu$m from the TSV. The impact of the oxide thickness of the TSV, T$_{oxtsv}$ is also investigated. A minimum thickness of 0.05 $\mu$m was considered. It cannot be reduced further for technological reasons since too a thin layer could create insulation problems. A maximum thickness investigated was not considered beyond 0.5 $\mu$m since the oxide thickness of the TSV should not be much thicker, for technological recommendations, in order to avoid the increase in the diameter of the TSV. In the range of T$_{SUB}$ values considered, the thickness of the TSV insulation oxide also has a significant influence on nearby devices, hence the increase of such a parameter serves to reduce the coupling of the substrate.

\begin{figure}[!t]
	\centering
	\includegraphics[width=0.55\textwidth, height=0.275\textheight]{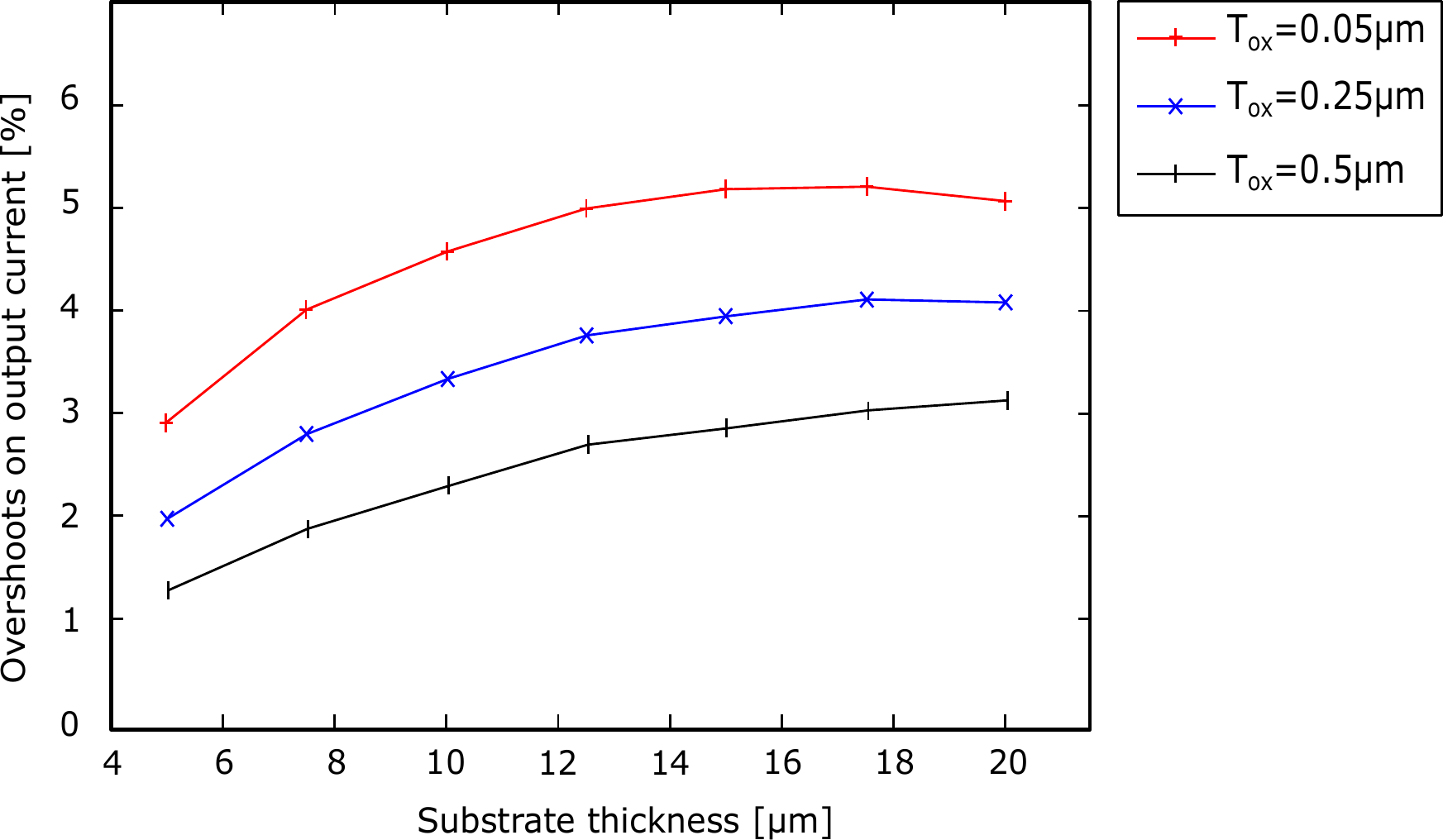}
	\caption{The overshoots observed on the drain current of the transistor versus the thickness of the substrate.
		\label{fig:fig3}}
\end{figure}

\begin{figure}[!t]
	\centering
	\includegraphics[width=0.55\textwidth, height=0.275\textheight]{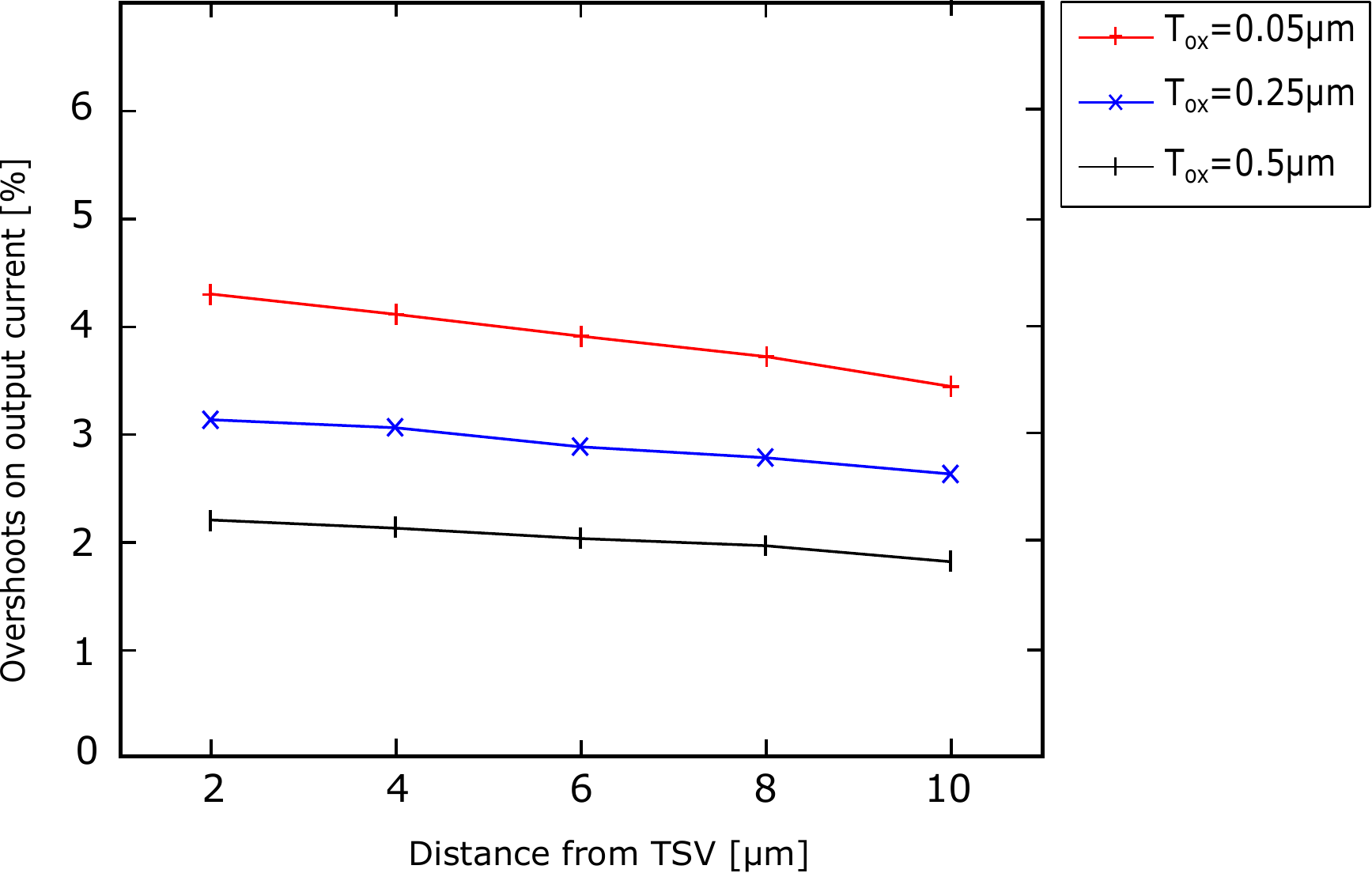}
	\caption{The overshoots observed on the drain current of the transistor versus the clearance distance from the via. 
		\label{fig:fig4}}
\end{figure}

\item
\textbf{The distance from the TSV:}
The separation distance of the TSV from the MOS transistor is an important design parameter, its influence is modeled for a range of distances that varies from 2 to 10 $\mu$m. These distances are maintained in this range of values which are not less than 2 $\mu$m, in order to take into account the constraint of misalignment of the TSVs with respect to neighboring devices, and not more than 10 $\mu$m in order to maintain a sufficiently high integration density. The influence of this parameter can be observed in Figure \ref{fig:fig4} which shows a considerable variation observed in the drain current of the MOS transistor as a function of this separation distance. The substrate thickness, in this case, is fixed at 10 $\mu$m and the oxide thickness of the via is also considered with the same values as before. However, from the results obtained, it is quite evident that the coupling of the substrate decreases when the separation distance increases, while it is higher at a minimum distance (2 $\mu$m). Also, in this case, the increase in the oxide thickness of the TSV helps reduce the parasitic coupling.
\item
\textbf{The TSV signal rising/falling time:}
 One of the electrical parameters investigated here is the rising and falling time of the signal applied to the TSV, $t_{rf}$.  In this study, the MOS transistor is placed close to the TSV with a 6 $\mu$m separation distance and a substrate thickness of  10 $\mu$m. Two values of oxide thickness of the TSV are considered in this case, with a minimum value of 0.05 $\mu$m and an average thickness of 0.25 $\mu$m. The study of the influence of this $t_{rf}$ parameter on the electrical characteristics of MOS components is done considering a range of values from 20 to 300 ps. The variations observed in the drain current of the MOS transistor as a function of this parameter are described in Figure \ref{fig:fig5}. One can note in this figure the influence of the rising/falling time of the signal applied to the TSV on the intensity of the coupling, it appears clearly that the short times generate a more consequent coupling compared to longer times.
\end{itemize}

\begin{figure}[!t]
	\centering
	\includegraphics[width=0.55\textwidth, height=0.275\textheight]{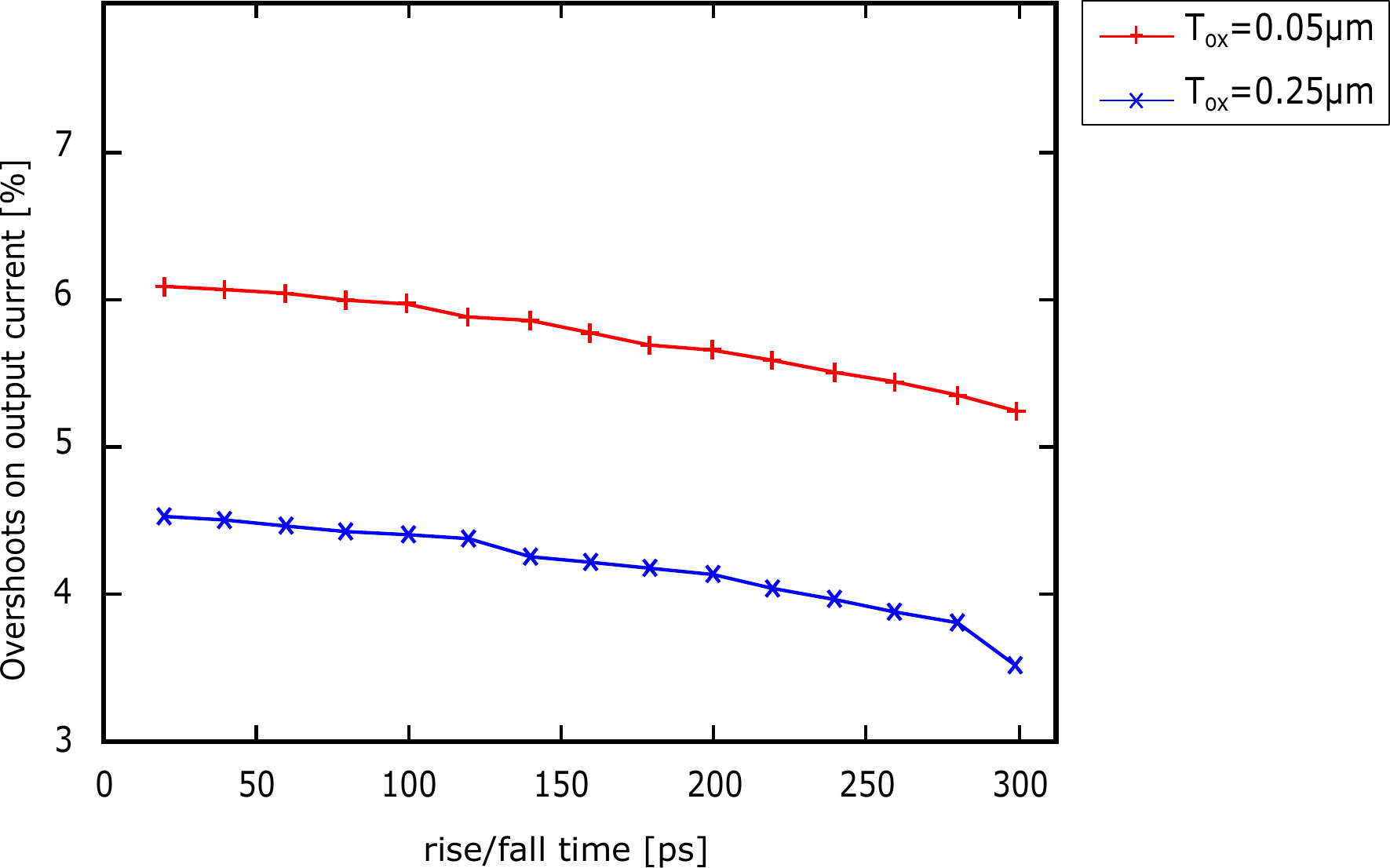}
	\caption{The overshoots on the drain current versus the TSV rising/falling time signal. 
		\label{fig:fig5}}
\end{figure}

\subsection{Analytical model}
When the accuracy of physics-based results is essential, numerical simulations would be the best solution to study the impact of 3D-TSV interconnects on the characteristics of CMOS ICs. However, when dealing with large circuits, the simulation is expected to take a long time. In this case, a SPICE-like circuit simulation is a viable option for obtaining quasi-instantaneous results \cite{jank}. This simulation requires the development of an analytical model which defines each element in the equivalent circuit to be investigated. 

\subsubsection{Device and analysis description}

This analytical model is defined in detail here showing the impact of TSVs on MOS transistors, which will be useful to extend the work to see its influence on CMOS circuits. The equivalent circuit is made up of a set of capacitors and resistors for the substrate and a set of MOS capacitors representing the via's insulation layer (an oxide layer). MOS transistors, on the other hand, will be modeled using a compact model (BSIM4).

\begin{itemize}

\item 
\textbf{Substrate network:}
\\

Silicon possesses dielectric and conductive properties, which can be translated into a capacitive and resistive effect, respectively. Therefore, the silicon substrate could be viewed and modeled whether as a purely resistive or capacitive and resistive network depending on the frequency used. Eq.1 defines this crossover frequency $f T$ as:\cite{ali}

\begin{equation}
f_{T} = \frac {1} {2\pi T_{s} } = \frac{q(p \mu_{p} + n \mu_{n} )} {2\pi \epsilon_{0} \epsilon_{si} }
\end{equation}

With $T_{s}$  the substrate dielectric relaxation time, q the electron charge, $\mu_n$ and $\mu_p$ stand for the electron and hole mobilities, respectively, and n and p are the carrier densities, and $ \epsilon_{si} $ the silicon permittivity.

Therefore, the equivalent circuit proposed describing the coupling through the substrate is made up of two main vertical ($R_{ver}$ and $C_{ver}$) and lateral ($R_{Lateral}$ and $C_{Lateral}$) components as shown in Figure \ref{fig:fig6}-a. These components present the substrate resistance and capacitance between the device's contacts and the TSV outer edge contact at a specific distance, $D_{TSV}$. Since this model is going to be validated later on using a linear geometry with a 2D-TCAD simulation, the cross-sectional area is set by default as 1 $\mu$m so that a square geometry is considered in this case. The expressions of equivalent resistances and capacitances for evaluating the parasitic coupling in the silicon substrate could be found in the literature as \cite{ali,raskin}:

\begin{figure}[!t]
	\centering
	\includegraphics[width=0.8\textwidth, height=0.275\textheight]{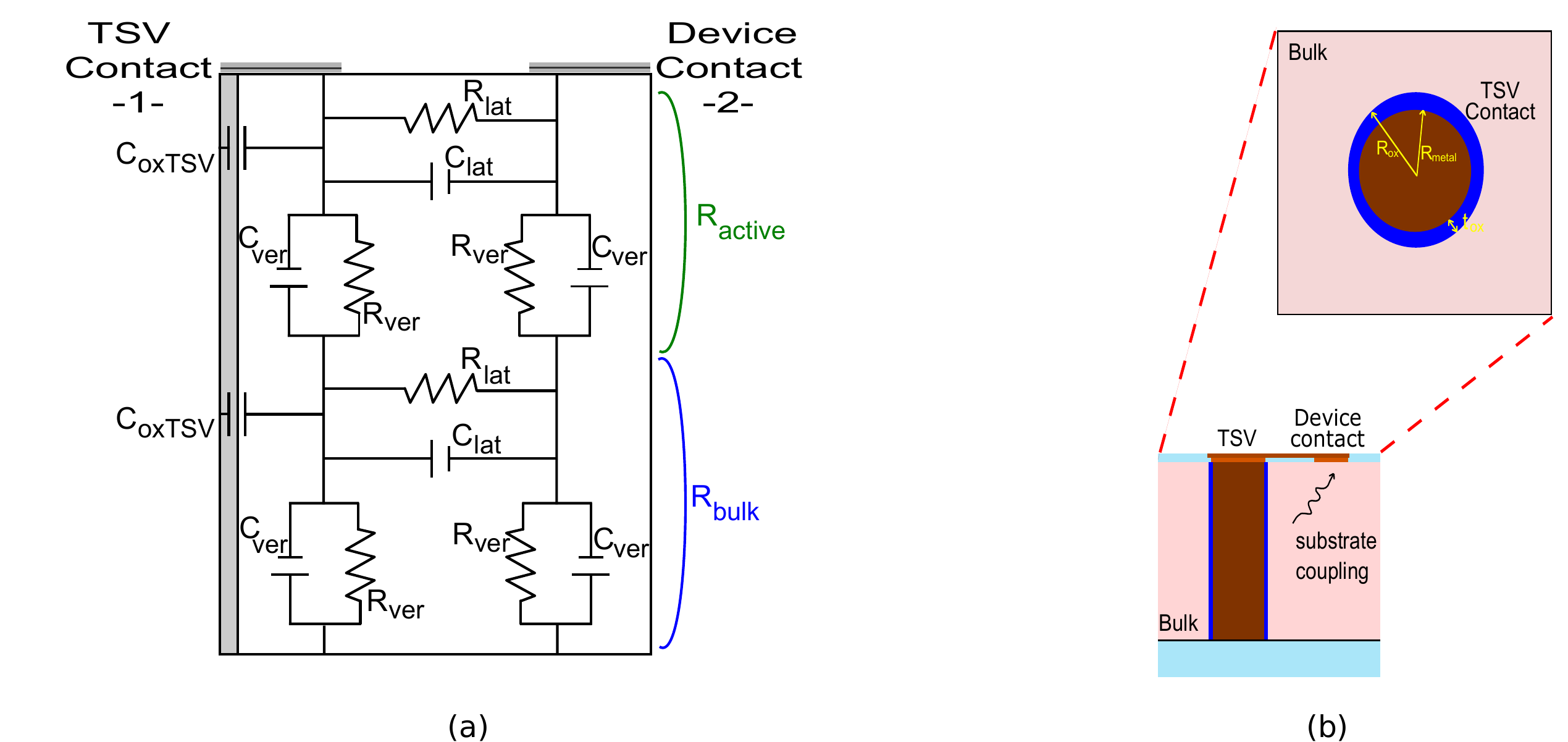}
	\caption{(a) Electrical model showing the different elements of the substrate network and TSV contact, and (b) schematic view of TSV 
                 for capacitor-related parameters. 
		\label{fig:fig6}}
\end{figure}

\begin{equation}
R_{ver} = [K_{1} \frac{ \sigma_{si} S_{pad}} {T_{sub} }] ^{-1}
\end{equation}

\begin{equation}
C_{ver} = K_{1} \frac{ \epsilon_{0} \epsilon_{si} S_{pad}} {T_{sub} }
\end{equation}

\begin{equation}
R_{Lateral} = [K_{2} \frac{ \pi \sigma_{si}} { 4ln [\frac{\pi (D_{TSV}-W)}{W+t}+1]}W] ^{-1}
\end{equation}

\begin{equation}
C_{Lateral} = K_{2} \frac{ \pi \epsilon_{0} (\epsilon_{si} +1)} {4ln [\frac{\pi (D_{TSV}-W)}{W+t}]}W
\end{equation}	

where $\sigma_{si}$ is the silicon conductivity, $ S_{pad} $=WxW is the pad surface, $t$ is the conductor thickness, $ K_{1} $ and $ K_{2} $ represent the fringing factors, that in turn can be defined as \cite{Rous,amine}:			

\begin{equation}
K_{1} = C_{e} \frac{ D_{TSV}} { \epsilon_{0} \epsilon_{si} S_{Pad}}
\end{equation}

\begin{equation}
K_{2} = (C_{o} - C_{e}) \frac{ 2ln [\frac{\pi (D_{TSV}-W)}{W+t}+1]} {\pi \epsilon_{0} (\epsilon_{si} +1) W}
\end{equation}

with $ C_{o} $ and $ C_{e} $ the equivalent capacitance of the even- and odd-modes for two open-end coupled microstrip lines expressed as:

\begin{equation}
C_{e,o} = \frac { \sqrt{\mu_{0} \epsilon_{0} \epsilon_{eff_{e,o}}}} {Z_{C_{e,o}}} W + C_{f_{e,o}}
\end{equation}

with $ \epsilon_{eff e,o} $ and $ Z_{C e,o} $ the effective permittivities and the characteristic impedances at the end of circuits,  respectively, $ C_{f e,o} $ the fringing capacitances for both even and odd-modes. The effective permittivities and characteristic impedance expressions could be defined in terms of other circuit geometries (such as substrate thickness, $T_{sub} $ and distance between via and the device, $ D_{TSV} $), which are used in the expressions illustrated in details in \cite{kir}. The same idea is used to define the fringing capacitance expressions from \cite{ed}.

Since the substrate network elements of both resistances and capacitances have values that are directly related to the local doping of the structure, the doping concentration is not constant throughout the substrate, the highly doped surface region, called 'active', is modeled as equivalent resistance $ R active $ and equivalent capacitance $C active$. Whereas, at the bottom of the structure, the doping is constant and is considered as a region called `bulk' modeled with an equivalent resistance $ R_{bulk} $ and capacitance $C_{bulk}$, as can be seen in Figure \ref{fig:fig6}-a. In our case study, the bulk region is considered with a high resistivity of 3.4 Ohm.cm while the heavily doped region with a low resistivity of 0.072 Ohm.cm \cite{Rous}. Such values are determined using an approximation of the average doping in both regions.

Values of the various components of the considered structure, with a TSV placed on the left edge, can be calculated using the analytical expressions expressed above as a function of different parameters like the substrate thickness and the distance between the TSV and the device contacts.

The capacitance values obtained from these expressions are very small (less than 0.2fF) and negligible, indicating that the substrate network is purely resistive. On the contrary, the resistance of the substrate network is strongly dependent on the geometry of the structure. As shown in Figure \ref{fig:fig7}-a, increasing the substrate thickness (TSV's length) increases the vertical $ R_{bulk} $ while keeping the rest of the resistances (lateral $ R_{bulk} $ and a$ R active $) almost constant. However, as shown in Figure \ref{fig:fig7}-b, increasing the separation distance between the via and the device contact increases both lateral $ R_{bulk} $ and $ R active $, that is much smaller than the $ R_{bulk} $ because of its low resistivity. As a result, the active region is regarded as the location for the parasitic substrate coupling, which must be reduced in this work.

\begin{figure}[!t]
	\centering
	\includegraphics[width=0.9\textwidth, height=0.3\textheight]{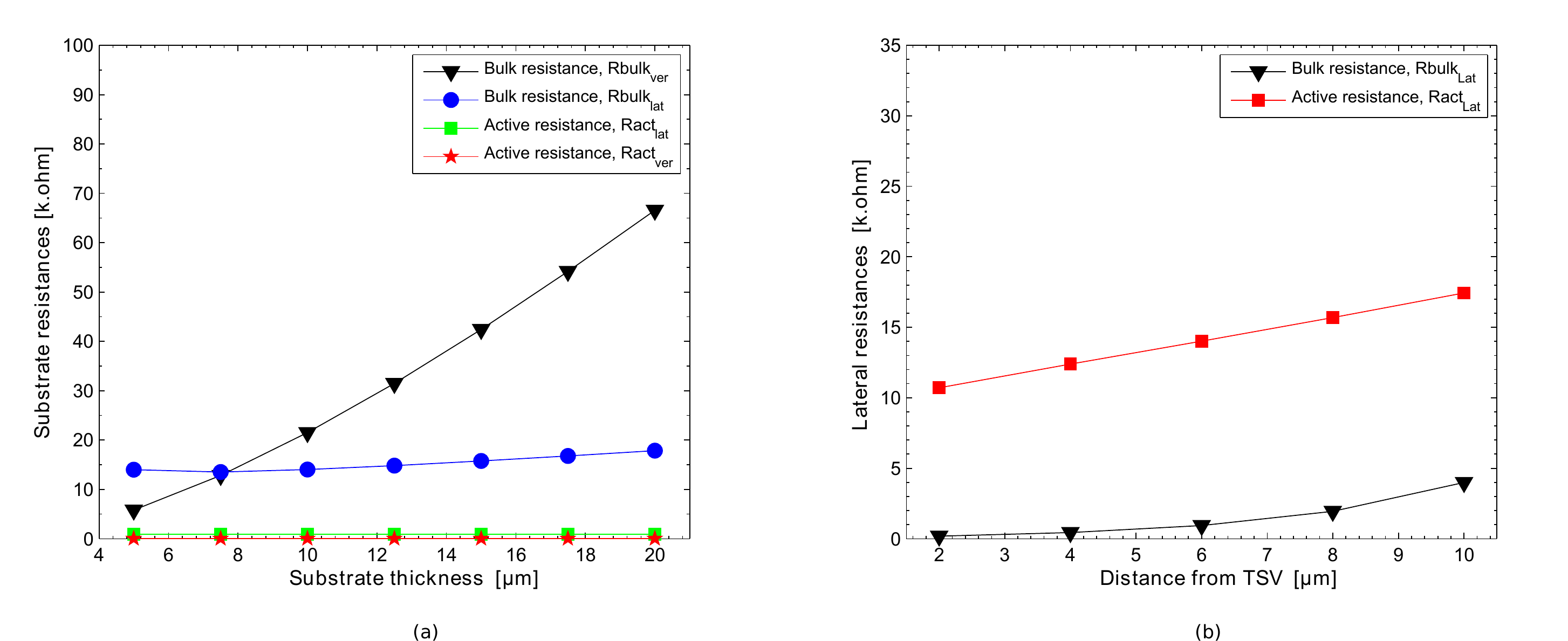}
	\caption{(a) Vertical and lateral bulk and active regions resistances versus the thickness of the substrate, and (b) Lateral bulk and active regions resistances versus the separation of the TSV contact from the device contact.
		\label{fig:fig7}}
\end{figure}

\item 
\textbf{TSV analytical model:}
\\
The following equation defines the oxide total capacitor of the TSV used in this model \cite{katti}:

\small
\begin{equation}
C_{ox} = \frac { 2 \pi \epsilon_{ox} T_{sub}}  {\ln (\frac{R_{ox}}{R_{metal}})}
\end{equation}
\normalsize

where $ R_{ox} $ is the radius of the TSV's oxide layer, $ R_{metal} $ the radius of the metal of the via, $ \epsilon_{ox} $ the oxide permittivity, $ T_{sub} $ the length of the Via or in other terms the substrate thickness (see Figure \ref{fig:fig6}-b. The TSV oxide layer is considered as a set of MOS capacitors distributed across the entire silicon depth covering both bulk active regions. The calculated total capacitance is a linear function of the substrate thickness, in which a small substrate thickness of 5 $\mu$m shows a capacitance of 22.2 fF and 4.85 fF for $T_{ox}$ = 0.05 $\mu$m and $T_{ox}$ = 0.25 $\mu$m, respectively. Whereas, with larger $ T_{sub} $ up to 20 $\mu$m, the capacitance goes up to 88.79 fF and 19.41 fF for $T_{ox}$ = 0.05 $\mu$m and $T_{ox}$ = 0.25 $\mu$m, respectively.

\end{itemize}

\subsubsection{Device description and analysis method}
\begin{figure}[!t]
	\centering
	\includegraphics[width=0.6\textwidth, height=0.25\textheight]{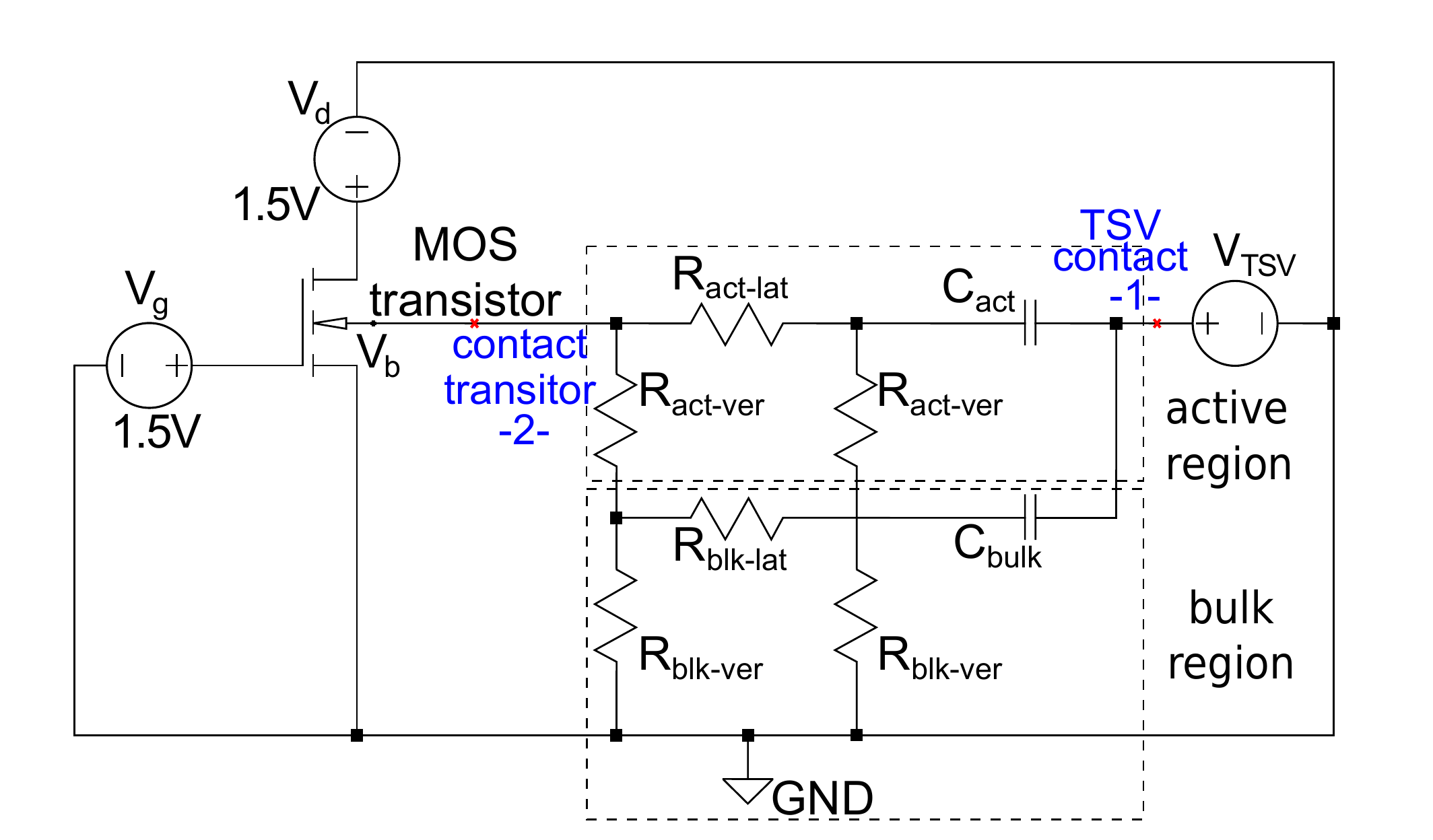}
	\caption{Circuit schematic used to study the 3D-TSV technology impact on a single MOS transistor.
		\label{fig:fig8}}
\end{figure}

A SPICE model was developed using the circuit analysis tool HSpice-Synopsys to investigate the effect of 3D-TSV interconnects on MOS transistors. This approach is best suited to perform comprehensive characterization, particularly when complicated circuits and several parameters involved are in question. The MOS transistors were modeled using a compact model BSIM4 with a 65 nm technology, a channel length of 50 nm, and a gate oxide layer of 3.2 nm thick. The effect of TSVs on MOS transistors is investigated against different technological and electrical parameters, including the thickness of the substrate ($T sub$), the distance between the via and MOS transistors contacts ($D TSV$), and the signal rising/falling time ($t rf$). Figure \ref{fig:fig8} depicts the electrical model schematic considered made up of the various circuit components determined previously.

\begin{figure}[!t]
	\centering
	\includegraphics[width=0.65\textwidth, height=0.275\textheight]{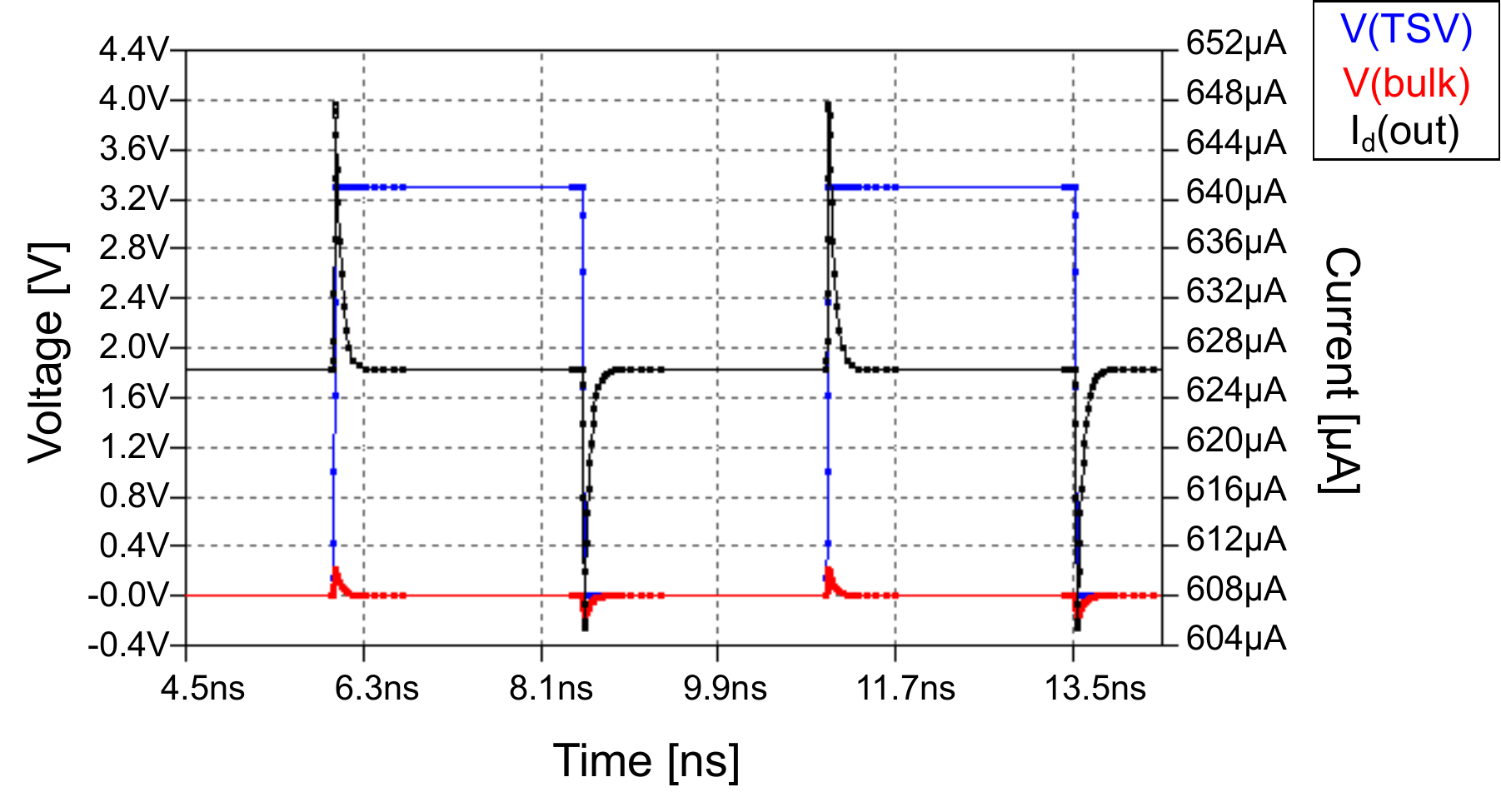}
	\caption{Simulated TSV interconnects influence on the electrical characteristics of a MOS transistor.
		\label{fig:fig9}}
\end{figure}

This proposed model is considered for both nMOS and pMOS transistors using specific compact models, where all quantities are complementary with equivalent bulk and active resistances. The signal applied to the TSV, defined as V$_{TSV}$, is considered with a 3.3 V amplitude of a square wave potential. The impact of the via on the characteristics of a single MOS transistor, regarding both bulk voltage and drain current, was investigated and is plotted in \ref{fig:fig9}.

\subsubsection{Parameter optimization}

The different parameters optimized are the same ones investigated previously, and which were performed using the analytical model proposed.

\begin{itemize}
\item 

\textbf{The substrate thickness:}
This first parameter, $T_{sub}$, was investigated in a range of values from 5 $\mu$m up to 20 $\mu$m. The overshoots percentage taken from the drain current of both nMOS and pMOS transistors versus the substrate thickness is presented in Figure \ref{fig:fig10}-a. These MOS transistors are placed at a 6 $\mu$m distance from the via. The impact of the oxide thickness of the TSV, $T_{ox}$ is shown also in this figure, with two different values. It can be seen that the substrate thickness has a clear impact on the performance of MOS transistors. The increase of the drain current depends on the threshold voltage, that in turn depends on the bulk voltage. This is explained by the increase in the vertical component of the bulk resistance network. Moreover, the induced negative coupling effects are higher in the case of a pMOS transistor in comparison to the nMOS transistor, because the Body effect is more pronounced. Within the interval of $T_{sub}$ values up to 10 $\mu$m, the oxide thickness of the TSV has a significant impact on transistors nearby, hence the increase of this parameter reduces the substrate coupling.

\begin{figure}[!t]
	\centering
	\includegraphics[width=0.9\textwidth, height=0.53 \textheight]{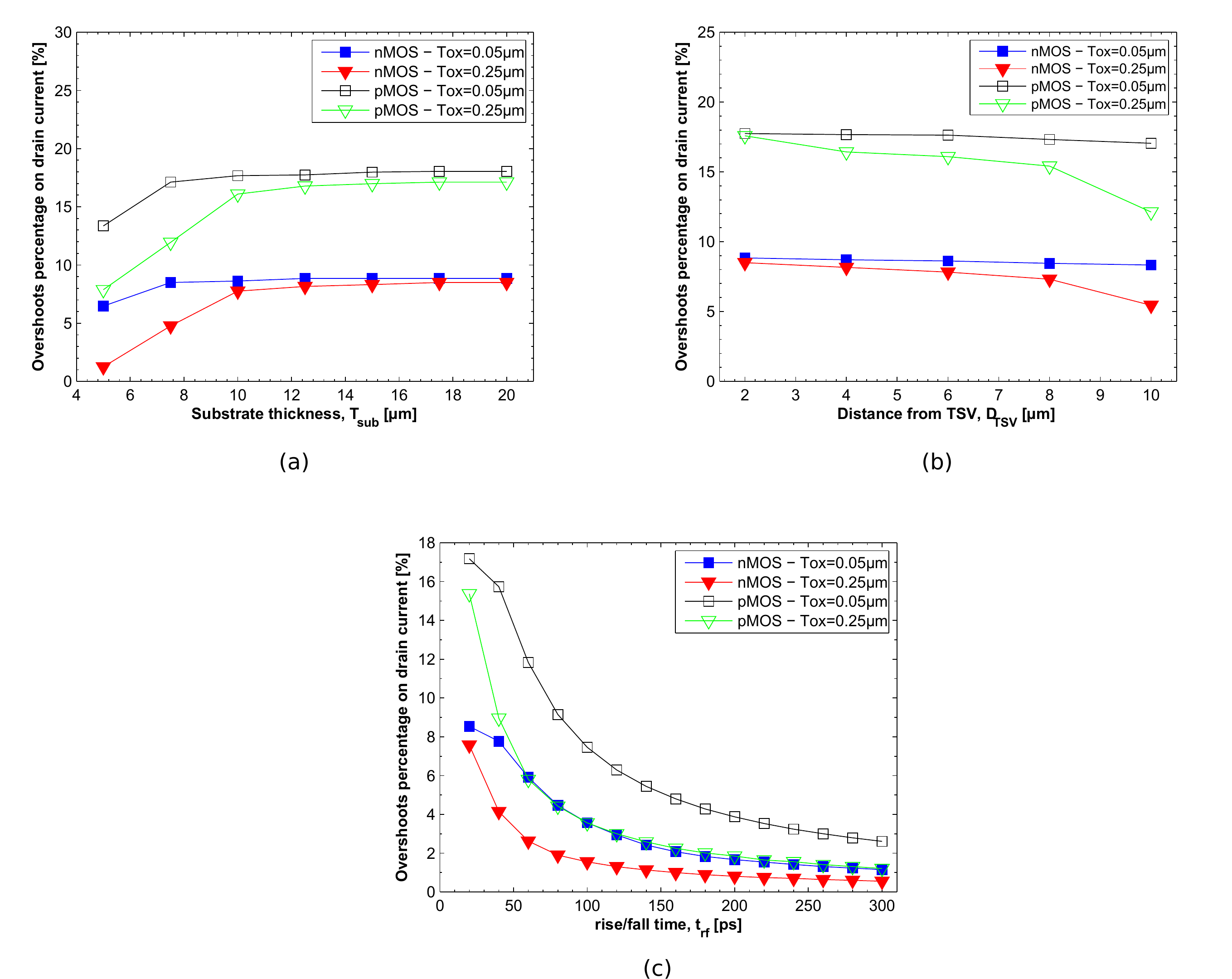}
	\caption{(a) The overshoots percentage taken from the drain current of MOS transistors versus the  
                substrate thickness, and (b) the overshoots percentage taken from the drain current versus the separation distance of the TSV from the MOS transistors contacts, and (c) the overshoots percentage taken from the drain current versus the rising/falling time signal.
		\label{fig:fig10}}
\end{figure}

\item 
\textbf{The distance from the TSV:}
The separation distance of the TSV from the device contact, $D_{TSV}$, was varied between 2-10 $\mu$m range. Its effect is seen clearly through Figure \ref{fig:fig10}-b. This figure shows the overshoot percentage taken from the drain current of the nMOS and pMOS transistors. The substrate thickness of these MOS transistors is fixed to 10 $\mu$m.  From this figure, it is evident that the substrate coupling decreases with the increases of $D_{TSV}$. This is due to the increase of both lateral $ R_{bulk} $ and $ R active $ resistances.  Increasing the TSV oxide thickness can also reduce the substrate coupling in this case.

\item 
\textbf{The TSV signal rising/falling time:}
This parameter presents the transition time of the TSV's square wave signal between two states, 1 and 0, and thus has a direct impact on the charge and discharge of the oxide capacitance of the TSV. Figure \ref{fig:fig10}-c shows the percentage of overshoots taken from the drain current of both MOS transistors versus the rising/falling time, $t rf$, which varies from 20 ps to 300 ps. MOS devices are considered with a 10 $mu$m thick substrate and placed at a distance of 6 $mu$m from the via contact. The plots and results obtained show that short $t rf$ values have a significant impact on substrate coupling, which is amplified in the case of thin oxide thickness.

\end{itemize}

Comparing results obtained from the numerical simulation previously, the variation observed was in the range of 4 to 8\%, whereas the analytical approach seems to overestimate the impact of TSV at small substrate thicknesses, however, this difference doesn't exceed more than 2\%. The good agreement with the obtained results from the  analytical model proposed confirms the possibility to use the proposed analytical approach to evaluate the 3D-TSV interconnects impact on the performance of CMOS circuits.

\section{3D-TSV influence on CMOS circuits}
The use of the proposed analytical approach was extended to evaluate in this case the influence of TSVs on the characteristics of CMOS ICs, and particularly in this work the CMOS inverter and the ring oscillator.

\subsection{CMOS inverter}
\begin{figure}[!b]
	\centering
	\includegraphics[width=0.6\textwidth, height=0.25 \textheight]{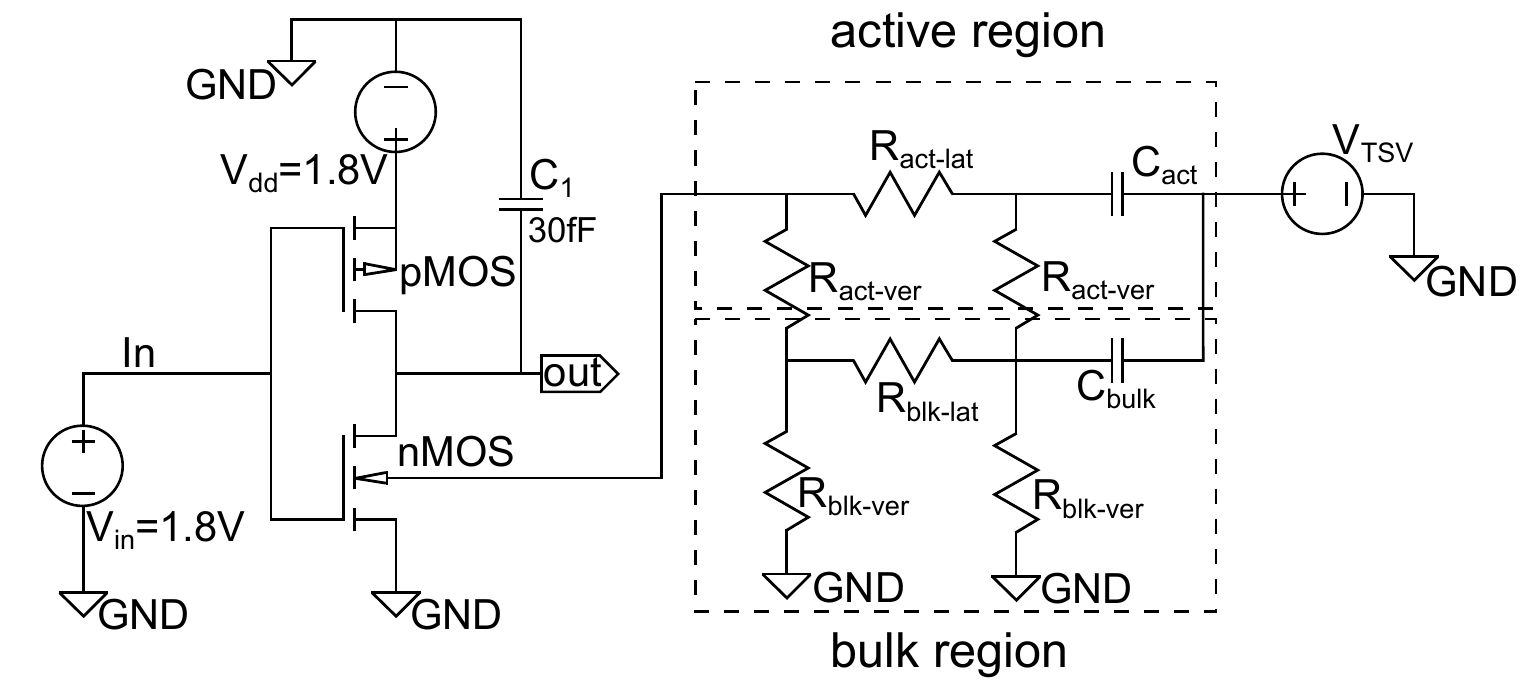}
	\caption{Schematic view of the SPICE circuit model for the study of the impact of TSV on a CMOS 
                 inverter circuit.
		\label{fig:fig11}}
\end{figure}

\begin{figure}[!b]
	\centering
	\includegraphics[width=0.65\textwidth, height=0.25 \textheight]{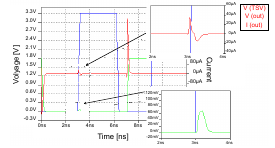}
	\caption{Impact of TSV on the CMOS inverter output current and output voltage.
		\label{fig:fig12}}
\end{figure}

The electric model for this analysis necessitates the use of both types of MOS transistors, as shown in Figure \ref{fig:fig11}. In this model, because the N-well acts as a shield toward the TSV-induced coupling preventing it from reaching the active area of the pMOS transistor, only the nMOS transistor is connected to the substrate resistance network. The TSV's influence is being investigated with respect to the same parameters considered previously. Applying a square wave signal on the TSV of a 3.3 V amplitude with a frequency of 200 MHz and a rise/fall time of 200 ps, the simulated plot is shown in Figure \ref{fig:fig12}. In this analysis, the $V DD$ of the CMOS inverter is set at 1.8 V while its input was biased with a square wave voltage of 1.8 V and a 72 MHz frequency \cite{Rous}.

\subsubsection{The substrate thickness}

\begin{figure}[!t]
	\centering
	\includegraphics[width=0.9\textwidth, height=0.53 \textheight]{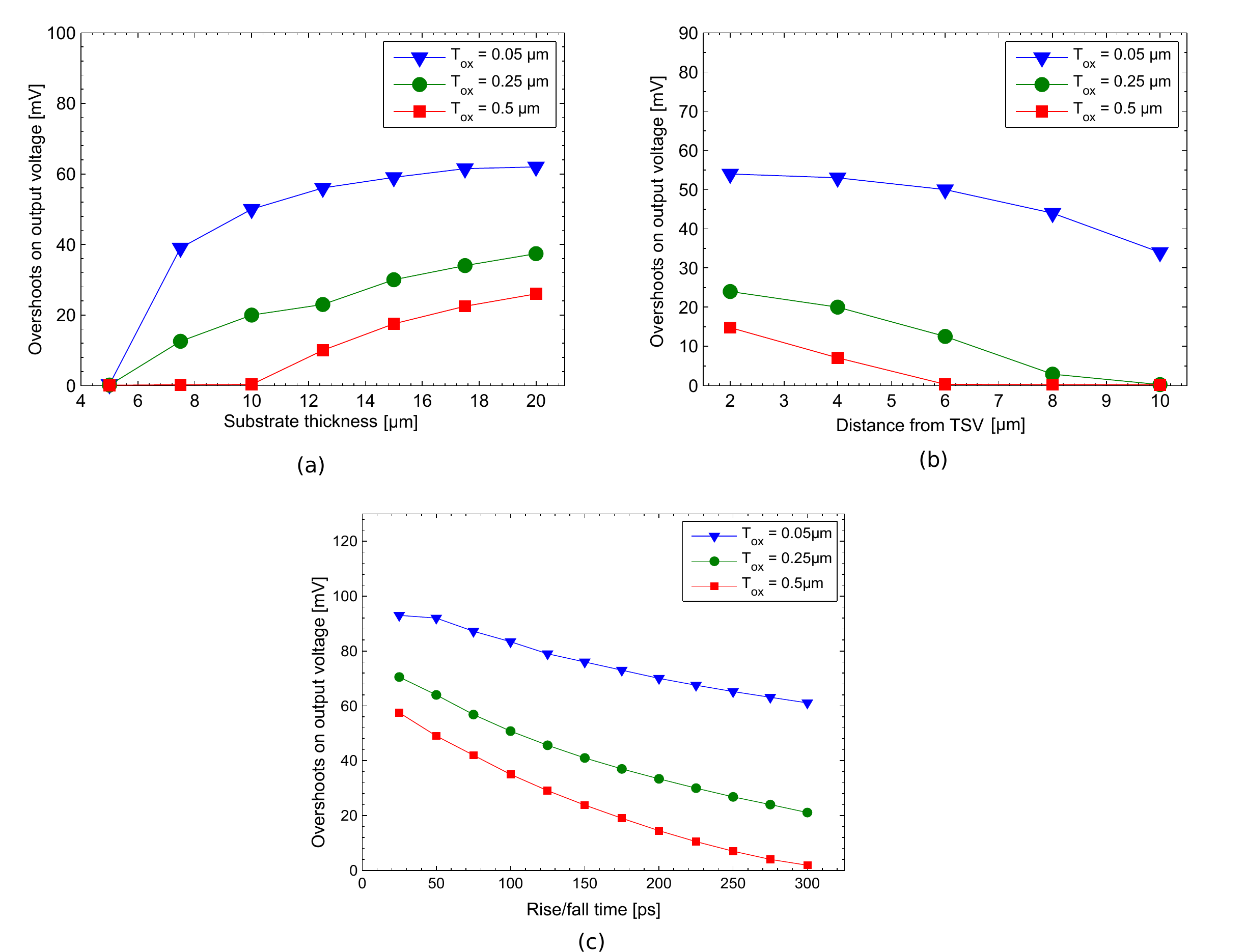}
	\caption{(a) The overshoots on the inverter output voltage as a function of the substrate thickness, and (b) 
                     The overshoots on the inverter output voltage as a function of the distance from the TSV contact, and (c) The overshoots on the inverter output voltage versus the TSV signal rising/falling time.
		\label{fig:fig13}}
\end{figure}

The influence of the 3D-TSV interconnect on the performance of the CMOS inverter is studied first with respect to the substrate thickness that was varied in a range of values from 5 $mu$m up to 20 $mu$m (see Figure \ref{fig:fig13}-a). The CMOS inverter parameter investigated concerns the output voltage. Placing the TSV at a 6 $mu$m distance from the inverter, the overshoots on its output voltage induced by the via increases as the thickness of the substrate increases as well, which is consistent with the effects observed on the nMOS transistor alone. 

\subsubsection{The distance from the TSV}

Figure \ref{fig:fig13}-b depicts the overshoots on the output voltage of the CMOS inverter versus the separation distance of the TSV from the inverter contact. This parameter was investigated from 2 $mu$m up to 10 $mu$m taking into account also different thickness values of the TSV's oxide layer.

It is noted through the simulation results that the effect of the TSV on the inverter output voltage is minimal once the separation distance is large, whereas it gets more significant the closer the via to the inverter circuit. This is explained by the fact that both lateral active region and bulk resistances value that increase with distance. on the other hand, it was observed that a thicker oxide layer of the TSV helps reduce the substrate coupling.

\subsubsection{The TSV signal rising/falling time}

Figure \ref{fig:fig13}-c depicts the overshoots on the output voltage of the inverter versus the via's signal rising/falling time. This parameter was investigated  With a range of 25- 300 ps, considering an inverter circuit placed at a separation distance of 6 $mu$m from the via and a substrate of 10 $mu$m thick. From the simulation results, it is observed that the overshoots on the output voltage of the inverter circuit decrease as $t rf$ get higher. This is mainly due to the effect of charge and discharge of the oxide layer capacitance of the via. 

Even though the simulation results show that the induced coupling by the 3D-TSV interconnects causes no logical errors in the inverter circuit performance, it is still not negligible at least regarding the induced power consumption on the nMOS transistor.

\subsection{Ring oscillator}
\begin{figure}[!t]
	\centering
	\includegraphics[width=0.9\textwidth, height=0.175 \textheight]{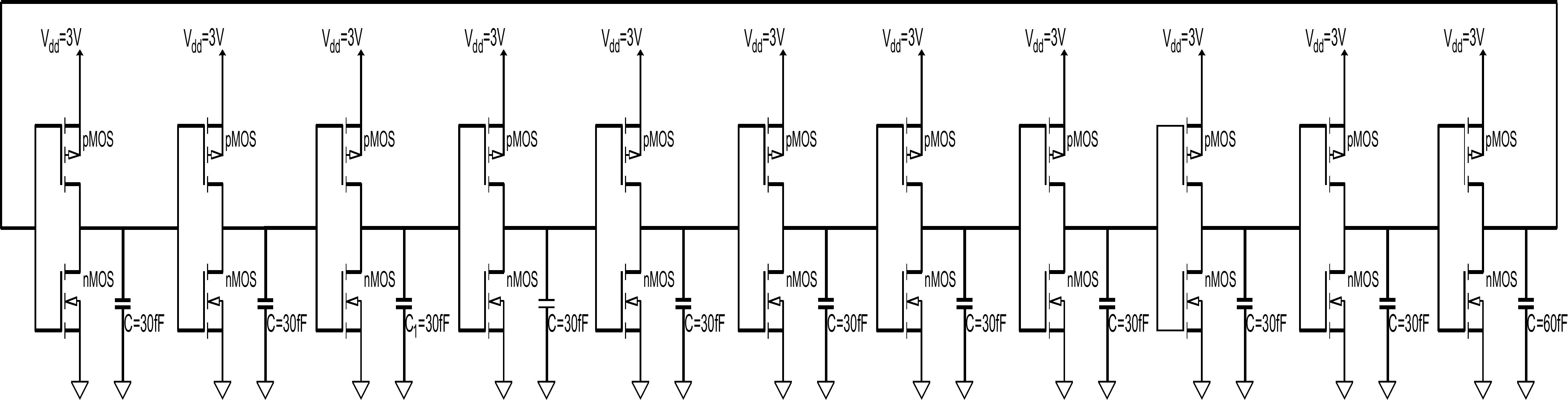}
	\caption{Schematic view of the 11-stage ring oscillator circuit modeled in SPICE.
		\label{fig:fig14}}
\end{figure}

\begin{figure}[!t]
	\centering
	\includegraphics[width=0.725\textwidth, height=0.3 \textheight]{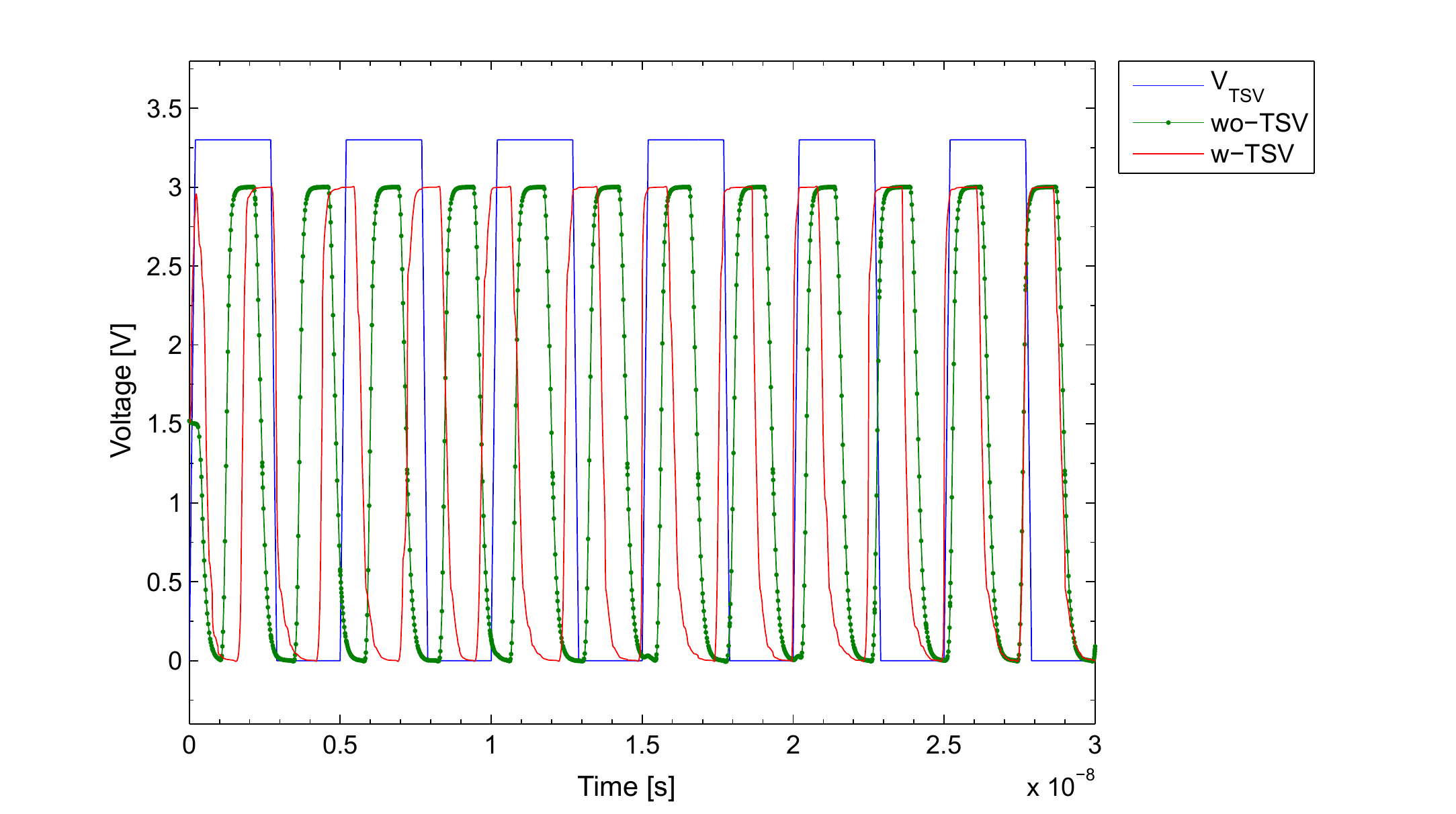}
	\caption{Simulated output voltage of the 11-stage ring oscillator with and without the presence of 3D-TSv interconnects nearby.
		\label{fig:fig15}}
\end{figure}

Based on the same approach considered to investigate the TSV-induced coupling in the inverter circuit, we have evaluated, in this section, the impact of the 3D-TSV technology on a Ring Oscillator. For this purpose, we have taken as a reference an 11-stage ring oscillator circuit without the presence of any via connections nearby (see \ref{fig:fig14}). Then, the electrical model to study the parasitic coupling induced by the TSV interconnects was implemented. The output signal of the 11-stage oscillator with a TSV contact nearby was simulated considering a TSV biasing with a square wave potential of 3.3 V amplitude, a frequency of 200 MHz, and a signal rise/fall time of 200 ps. The oscillator simulation results were plotted in Figure \ref{fig:fig15} where the 11 inverters were biased with $V_{DD}$ of 3 V.         

As a figure of merit for the ring oscillator, the delay time per gate is considered, that is defined as: 

\begin{equation}
\tau = T/2n 
\end{equation}	

when T is the oscillator time period and n the number of its gates (n=11 in the considered example).

\subsubsection{The substrate thickness}
The time delay variation of the ring oscillator induced by the TSV was investigated versus the substrate thickness. This parameter was varied in a range of values from 5 $mu$m up to 20 $mu$m and the oscillator circuit is placed at 6 $mu$m away from the TSV contact.   

From the simulation results shown in Figure \ref{fig:fig16}-a, it is evident that increasing the thickness of the substrate has a strong contribution and influence on the time delay of the circuit. Such behavior is much evident once a thinner TSV oxide layer is considered, as already seen in previous studies.  

\subsubsection{The distance from the TSV}	
\begin{figure}[!t]
	\centering
	\includegraphics[width=0.9\textwidth, height=0.53 \textheight]{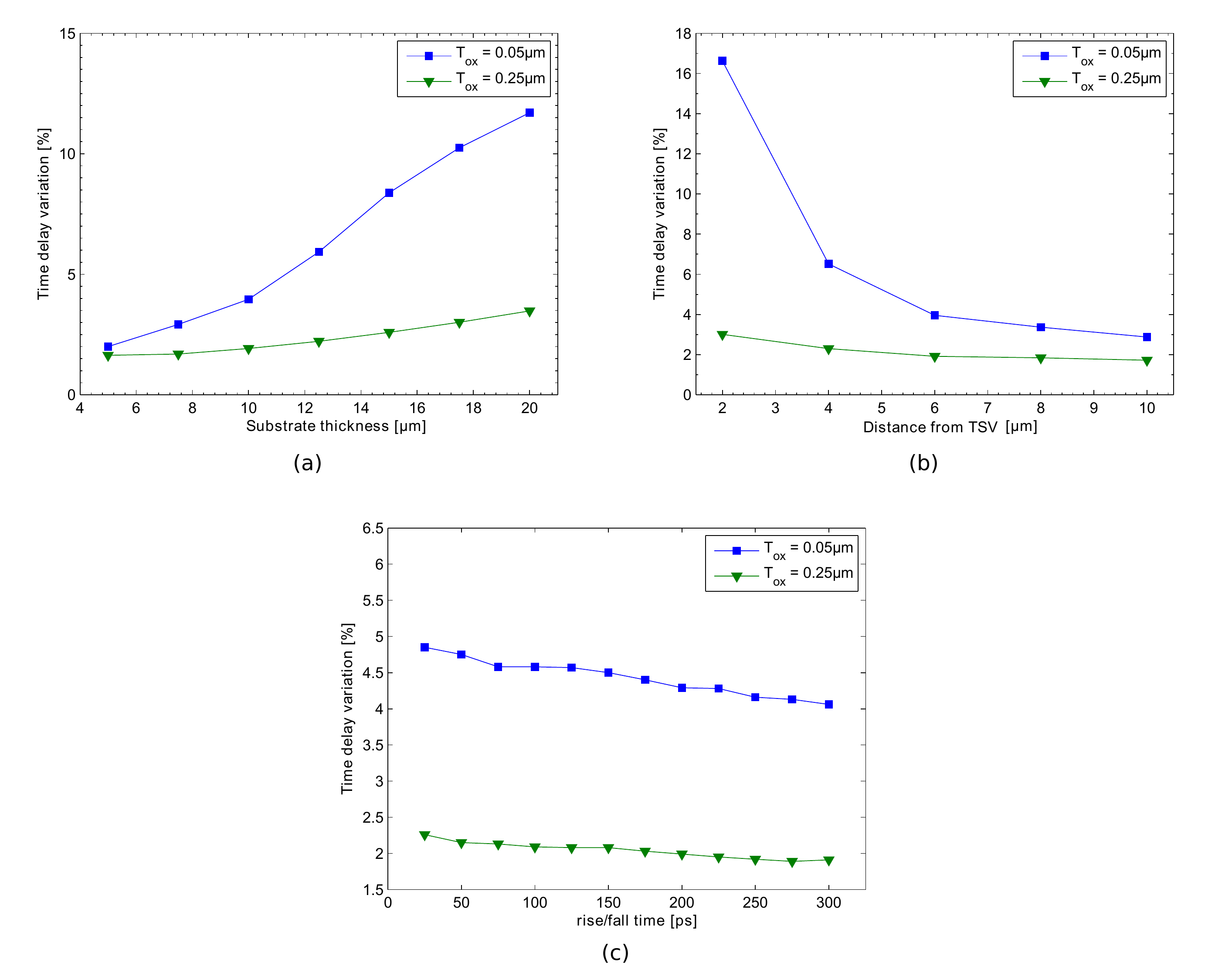}
	\caption{Variation observed on the time delay of a ring oscillator circuit with respect to : (a) substrate thickness, (b) TSV separation distance, and (c) the signal rising/falling time.
		\label{fig:fig16}}
\end{figure}

This technological parameter has a clear impact on the performance of the ring oscillator, as illustrated in Figure \ref{fig:fig16}-b. This influence is expressed by the variation in time delay simulated versus such separation distance ranging from 2 $mu$m to 10 $mu$m.

From the simulation results obtained, it is clear that the TSV-induced time delay of the oscillator circuit is considerable once the separation distance is a few micrometers, whereas it gets lower at large distances. This is still explained by the fact that the increase in both lateral bulk and active region resistances causes a decrease in the parasitic coupling from the via.

\subsubsection{The TSV signal rising/falling time}	
Figure \ref{fig:fig16}-c depicts the TSV signal rising/falling time influence on the ring oscillator. This parameter was varied in a range of 25-300 ps and simulation results show that it does not show a considerable influence on the time delay of the oscillator in comparison to the previous cases. 

The TSV has a greater impact on ring oscillator performance than on inverter performance, causing significant variance in the performance of a simple structure of an 11-stage oscillator circuit. The observed impacts would be amplified in circuits with a higher number of stages.

\section{Conclusion}
In this work we reported on the evaluation of the impact of TSV interconnects on the electrical performance of MOS devices by means of numerical TCAD simulation and an analytical model once considering CMOS circuits. Both approaches were defined and implemented to describe the TSV behavior at the circuit level and other phenomena such as the substrate coupling, as a function of different technological and electrical parameters. The analytical model compared to the numerical approach has proved to be effective to perform rapid and reliable simulations of the substrate coupling induced by the TSV on two commonly used CMOS circuits, i.e., the CMOS inverter and the ring oscillator. Range of values for which the impact of the considered parameters (i.e., the substrate thickness, the distance from the TSV, and the signal rising/falling time) on the circuit performance is limited 
can be easily appreciated from the simulations, thus yielding a useful design tool for use with even more complex CMOS circuits. 

\bibliographystyle{unsrtnat}
\bibliography{references}  

\end{document}